%% file: paper.tex
\documentclass[10pt,sigconf,letterpaper,nonacm]{acmart}

\usepackage{xurl}
\usepackage{enumitem}
\usepackage{hyperref}
\usepackage{xcolor}
\usepackage{tcolorbox}
\usepackage{subcaption}

\renewcommand\footnotetextcopyrightpermission[1]{} %
\newcommand{\tool}{\textit{VPN-Scope}}
\newcommand{\cloudlab}{CloudLab}

\newtcolorbox{boxx}[1][]{
  colback=yellow!3,
  colframe=black!80,
  top=5pt,
  bottom=5pt,
  left=5pt,
  right=5pt,
  boxsep=0pt,
  before skip=4pt,
  after skip=4pt,
  #1
}

\setcopyright{none}

\acmDOI{}

\acmISBN{}

\acmConference[Submitted for review to ACM IMC]{}
\acmYear{2026}
\copyrightyear{}

\acmPrice{}

\begin{document}

\title{Not All Roads Lead to Rome:}
\subtitle{How VPN Selection Alters What We Measure and Infer about Web Infrastructure}
\renewcommand{\shorttitle}{Not All Roads Lead to Rome}


\author{Sachin Kumar Singh}
\affiliation{\institution{University of Utah}\city{Salt Lake City}\state{UT}\country{USA}}

\author{Robert Ricci}
\affiliation{\institution{University of Utah}\city{Salt Lake City}\state{UT}\country{USA}}

\author{Alexander Gamero-Garrido}
\affiliation{\institution{University of California, Davis}\city{Davis}\state{CA}\country{USA}}

\begin{abstract}

Web-measurement studies treat commercial VPNs as interchangeable vantage points within a country, assuming that any VPN in a particular country is as good as any other. 
We show that this assumption does not hold: the same country measured through different VPN providers yields materially different conclusions about where endpoints sit, who hosts them, and which physical replicas serve them. 
Using large-scale browser-based measurements across fourteen countries and four major VPN providers, complemented by targeted DNS and replica-selection probes, we examine sources of this variability across three layers of the VPN-to-endpoint path: vantage identity, name resolution, and replica selection. 
We find that the variability is driven primarily by layers below the client: commercial VPN providers operate their own in-country DNS infrastructure, often intercepting queries regardless of client configuration; CDNs steer on the exit network, sending identical queries to different replicas; and peering paths route identical DNS answers to different physical facilities. 
We distill these findings into a set of reporting practices for VPN-based Web measurement.

\end{abstract}

\maketitle

\input{body}

\bibliographystyle{ACM-Reference-Format}
\bibliography{reference}
\input{re-struc/appendix}

\end{document}

%% file: body.tex
\input{introduction}

\input{relatedwork}
\input{design}

\input{re-struc/method}

\input{re-struc/data}

\input{re-struc/rq1}

\input{re-struc/rq2}

\input{re-struc/rq3}

\input{discussion}
\input{conclusion}

%% file: introduction.tex
\section{Introduction}
The Web began as a relatively simple platform, but has since evolved into a complex, globally distributed, policy-constrained system~\cite{webalmanac2024-pageweight,webalmanac2024-cdn,EU2016GDPR,EU2022DSA}. 
It is now central to how people work, communicate, learn, and access essential services. 
To study this system, Web researchers often rely on systematic measurement to quantify performance and user experience~\cite{wang2013demystifying,singh2018characterizing,crux,kakhki2017taking} and to assess reliability and resilience~\cite{deccio2010measuring,singh2018characterizing,haq2022no,richter2018advancing,guillot2019chocolatine,tschantz2018bestiary,mcdonald2018403}. 
Measurements also help characterize the Web's infrastructure layer: mapping dependencies on third parties and CDNs~\cite{kashaf2020analyzing,RKumar-SIGMETRICS23,kumar2024choices,doan2022empirical,libert2015exposing}, tracing how user data flows across jurisdictions~\cite{gamero2025localization,singh2025world,iordanou2018tracing}, and grounding debates about data localization and regulation~\cite{degeling2018we,gamero2025localization,singh2025world}.

Many Web measurement studies rely on geographically distributed vantage points to capture location-specific behaviors~\cite{RKumar-SIGMETRICS23,kumar2024choices,gamero2025localization}. 
These vantage points are typically deployed through VPN services~\cite{RKumar-SIGMETRICS23,kumar2024choices,nobori2014vpn,khan2018empirical}, 
residential proxy networks~\cite{gamero2025localization,chung2016tunneling,chhabra2021measuring,choi2020understanding}, or cloud infrastructure~\cite{jueckstock2019blind,mok2021measuring}, each offering distinct trade-offs in terms of vantage points and measurement possibilities.
Among these, commercial VPNs are especially attractive and are widely used in prior work~\cite{RKumar-SIGMETRICS23,kumar2024choices,nobori2014vpn,khan2018empirical,niaki2020iclab,hoang2019measuring,jin2021understanding,eijk2019impact,roth2022security}, offering easy-to-acquire vantage points at low cost with broad country coverage.

However, VPNs are not transparent channels. VPN pro\-viders choose Points of Presence (PoPs), hosting networks, DNS resolvers, and peering arrangements, all of which can alter where requests appear to originate, how names resolve, and which mirrors or CDNs respond. 
If provider choices shape what we observe, VPN-based measurements may diverge from provider-neutral ones in ways that affect infrastructure level inferences downstream studies rely on: which countries endpoints sit in (cross-border exposure and data localization), which organizations host them (third-party dependency and centralization), and which physical replicas serve them (latency, reachability, and observed jurisdiction).
These inferences increasingly inform policy and regulatory debates, yet rest on VPN-based observations whose sensitivity to the VPN provider itself has not been characterized.

Despite the important implications of platform-induced variability, and while prior work has examined bias in other measurement platforms~\cite{bajpai2015lessons,sermpezis2023bias}, the research community still lacks a systematic understanding of the platform-induced variability introduced specifically by commercial VPNs in web measurements and, in turn, in the infrastructure inferences drawn from those measurements.
Closing this knowledge gap is challenging because providers disclose little about their operations and there is no community-standard dataset or toolkit aimed at evaluating VPN-induced variability for web measurement.

Our approach starts from a simple observation: if commercial VPNs were neutral conduits within a country, web measurements collected through different providers from the same source country should agree on the endpoints contacted, the organizations hosting them, and the replicas serving them.
Systematic disagreement is therefore evidence that the VPN itself is shaping the observation, and the structure of that disagreement points to where the variability comes from.
To that end, we build \textbf{\tool{}}, a web measurement tool that performs synchronized, repeated crawls across multiple VPN providers and countries, collecting HTTP Archive (HAR) traces of each page load for downstream analysis.
Using \textbf{\tool{}}, we collect 510K website snapshots across 14 geographically diverse countries and four major VPN providers, with multiple runs per country. 
To investigate the sources of the variability we observe, we complement this browser-driven dataset with three controlled measurements across the same vantages: a DNS-level experiment on 50K domains, a probe that reveals where DNS queries actually travel inside each commercial VPN tunnel, and an HTTPS-based probe that identifies which physical CDN replica each vantage reaches for a given anycast IP.

Our study is the first systematic attempt \textit{to investigate the variability introduced by VPN usage in web measurement data, to identify its structural sources, and to characterize how it shapes infrastructure-level inferences about the Web}.
The objective is not to evaluate or attribute specific effects to individual VPN providers, but rather to investigate whether VPN choice introduces measurable divergence in what we observe, why it does so, and how researchers should interpret measurements that depend on it.

Leveraging \textbf{\tool{}} and our complementary DNS measurements, we investigate three research questions:
\begin{itemize}[leftmargin=2em,itemsep=0pt,topsep=0pt]

        \item \textbf{RQ1 (Endpoint IPs and countries): }Within a given country, does VPN choice affect which IPs domains resolve to and the distribution of destination countries for contacted endpoints? (\S\ref{sec:rq1})

	\item \textbf{RQ2 (Hosting organization): }Do these differences persist at the organization level, or are they largely absorbed when IPs are mapped to their owners? (\S\ref{sec:rq2})

        \item \textbf{RQ3 (Sources of the variability):} Which layers of the VPN-to-server path---vantage location, DNS  resolution, CDN steering, and peering to anycast replicas---drive variability, and to what extent is each layer under a measurement study's control? (\S\ref{sec:rq3})

\end{itemize}

We find that
    \textbf{(RQ1)} VPN vantages are not provider-neutral, measurably shifting both the IPs returned for a domain and the destination country those IPs geolocate to;
    \textbf{(RQ2)} the divergence is largely absorbed at the organization layer, preserving inferences about hosting providers;
    \textbf{(RQ3)} the IP- and country-level variability is driven by in-country DNS infrastructure (which some VPNs intercept), and peering paths to anycast replicas---factors that measurement studies have little visibility or control over.
These effects are not uniform: significant provider-country interactions mean a provider's behavior in one region does not generalize, but the variability itself is persistent across runs and statistically detectable.

    We describe \tool{} in \S\ref{sec:design} and our data collection in \S\ref{sec:method}.
After examining each RQ (\S\ref{sec:rq1}--\S\ref{sec:rq3}), we discuss implications for the interpretation of past VPN-based
        studies in \S\ref{sec:disc-past} and make recommendations for future studies in \S\ref{sec:disc-future}.

%% file: relatedwork.tex
\section{Background and Related Work}

\textbf{Measurement vantage points.}
To understand the web at scale, researchers rely on public measurement platforms~\cite{ripeatlas,measurementlab,planeteu,caidaark,nlnogring}, commercial proxy services~\cite{brightdata,oxylabs}, self-deployed vantage points via cloud infrastructure~\cite{lee2021tls,bermudez2013exploring,afroz2018exploring,barron2017picky,jueckstock2019blind,mok2021measuring}, residential proxies~\cite{gamero2025localization,mi2019resident,yang2022extensive}, and virtual private networks (VPNs)~\cite{kumar2024choices,RKumar-SIGMETRICS23,maghsoudlou2023characterizing,xing2024yesterday,SNITCH-MADWEB25,CalcuLatency-SEC24,nobori2014vpn,khan2018empirical}.

\textbf{Residential proxies.}
Residential proxies (e.g., BrightData~\cite{brightdata}, Oxylabs~\cite{oxylabs}) are widely used to capture geographic and network diversity at scale. Using these proxies, prior work has audited EU data-localization compliance~\cite{gamero2025localization}, evaluated global DoH performance and geographic disparities~\cite{chhabra2021measuring}, revealed widespread TTL extension by ISP and mobile-network resolvers with cache and security implications~\cite{bhowmick2024measuring}, and characterized CDN anycast ``flipping'' that degrades page-load times~\cite{zhang2025characterizing}.

\textbf{Cloud vantage points.}
Cloud providers offer another class of distributed vantage points that is operationally scalable. Using AWS instances across multiple regions, prior work has measured TLS adoption and handshake performance, revealing uneven deployment and fallback behaviors~\cite{lee2021tls}, and has also studied global content access and on-path traffic manipulation~\cite{xing2024yesterday}. However, websites often treat cloud-hosted clients differently, limiting their value as general-purpose measurement nodes~\cite{afroz2018exploring}.

\textbf{VPN-based web measurement.}
VPNs have emerged as a widely used alternative for simulating region-specific access, valued for their ease of setup, commercial availability, and ability to support both browser- and network-level measurements.
While commercial proxies often operate at the application layer and are constrained to HTTP-level interaction, VPNs expose the full IP stack, enabling richer experimental designs.
Prior work has leveraged commercial VPNs as a convenient way to obtain geographically distributed vantage points~\cite{nobori2014vpn,khan2018empirical}, and more recent studies rely on VPN infrastructure to draw infrastructure-level inferences at scale: deploying VPN vantage points across 50+ countries to measure third-party service dependencies and centralization~\cite{RKumar-SIGMETRICS23}, examining government-website hosting across 61 countries and the jurisdictions in which that hosting sits~\cite{kumar2024choices}, measuring CDN geo-blocking and reachability at country granularity~\cite{mcdonald2018403}, and auditing EU data-localization compliance using VPN-assisted vantage points~\cite{gamero2025localization}.
Despite these advantages, VPNs are not neutral infrastructure: VPN providers' practices are opaque and may vary across regions, introducing artifacts that can bias measurements and yield inconsistent observations of web infrastructure and hosting composition.

\textbf{Studying the VPN ecosystem itself.}
A parallel body of work has examined the VPN ecosystem as an object of study rather than as a methodological lens. This thread has empirically analyzed commercial VPN providers~\cite{khan2018empirical}, systematically probed the VPN ecosystem for misconfigurations and leaks~\cite{ramesh2022vpnalyzer}, verified the physical locations of proxies and VPN exits~\cite{weinberg2018catch}, characterized the broader VPN ecosystem in the wild~\cite{maghsoudlou2023characterizing}, and leveraged latency-based techniques to detect proxy-enabled abuse~\cite{CalcuLatency-SEC24,SNITCH-MADWEB25}.

\textbf{How network vantage shapes what a client reaches.}
Closer to the mechanisms we isolate, prior work has documented how the client's network position shapes what it observes: global DNS manipulation at the ISP level~\cite{pearce2017global}, CDN anycast behavior measured through distributed vantages~\cite{sommese2020manycast2,cicalese2015characterizing}, the role of EDNS Client Subnet in CDN steering decisions~\cite{rfc7871}, and consolidation of the web onto a small number of infrastructure providers~\cite{doan2022empirical,kashaf2020analyzing}. Ultimately, VPN-based infrastructure studies treat VPNs as neutral vantage points, but provider selection and behavior can influence observations and, consequently, the infrastructure-level conclusions drawn from them.

\textbf{Our work.}
This paper directly examines the assumption that VPNs act as neutral, interchangeable vantage points in web-infrastructure measurement.
Prior work has not systematically asked how VPN \emph{choice} shapes the infrastructure inferences that VPN-based measurements produce---which countries endpoints sit in, which organizations host them, and which physical replicas actually serve them.
We show that VPN selection affects these infrastructure observables even when location is held constant, and we isolate the network-layer mechanisms responsible so that the effect can be interpreted. Rather than evaluating provider ``quality,'' our goal is to characterize the extent, structure, and origins of this variability across providers and geographies, laying the foundation for more robust measurement practice.

%% file: design.tex
\section{\tool{} Design}
\label{sec:design}

We designed and built \tool{} to answer questions regarding the ways that different VPNs may produce different outcomes when conducting measurements of websites.

\tool{} is a measurement framework that coordinates synchronized web measurements across VPN vantage points, collecting 
the data needed to assess how VPN choice affects web measurement observations and the infrastructure-level inferences drawn from them.
The design is guided by three goals:
(i) \textbf{isolation} between concurrent VPN sessions to avoid cross-contamination of routes, DNS state,
and browser state; (ii) \textbf{repeatability} via controlled, comparable execution across providers and regions;
(iii) \textbf{scalability} to support experiments across $n$ VPN configurations without dedicating one host per VPN.

\subsection{System overview}
\tool{} consists of three components:
\begin{enumerate}[leftmargin=*, noitemsep, topsep=2pt]
    \item \textbf{Vantage Manager}, which instantiates a per-VPN isolated network context, attaches a VPN tunnel
    within that context, and verifies connectivity and tunnel readiness.
   \item \textbf{Coordinator}, which schedules measurement tasks across VPN configurations and target websites, and launches concurrent (and optionally coordinated) measurement runs.

    \item \textbf{Measurement Runner}, which executes web measurements inside a selected vantage point by launching
    Selenium-driven browsing sessions for a website and recording HTTP Archive (HAR) trace.
\end{enumerate}

Although we instantiate vantages using commercial VPNs in this study, \tool{}'s namespace-based abstraction generalizes to other network contexts (e.g., distinct physical uplinks) by attaching each context to a dedicated namespace and executing the same measurement runner unchanged.

\subsection{Vantage Isolation}
\label{subsec:namespace}
\tool{} runs each VPN configuration inside its own Linux \emph{network namespace}. A namespace provides an isolated
network stack (interfaces, routes, and DNS settings), which prevents interference between concurrent VPN sessions and
allows many vantages to coexist on a single host. Network namespaces provide a practical balance between isolation and scalability, making it feasible to compare $n$
VPN vantage points under controlled execution and quantify how VPN choice influences the infrastructure-level observations a web measurement produces.

For each vantage, \tool{} creates a link between the host and the namespace to provide connectivity,
then establishes the local endpoint of the VPN tunnel \emph{within} the namespace. As a result, VPN-provisioned interfaces and routes are confined
to that vantage and do not affect the host or other concurrent experiments.
\tool{} configures a namespace-specific DNS resolver so that name resolution is scoped to the namespace and does not leak across concurrent runs. 
If a VPN client provides (or overrides) DNS settings when the tunnel is established, those changes apply only within that namespace.
After a namespace is instantiated, \tool{} launches the VPN client inside that namespace so that the tunnel interface and any
VPN-installed routes remain local to the vantage. Simultaneous vantages can be established for the same or different VPN providers, different remote endpoints, etc. enabling efficient experimentation. In the deployment used for this paper, we select a destination country, establish vantages for each of the commercial VPNs, run all experiments for that country, then create new vantages for the next country.

Before executing any web measurements, \tool{} performs lightweight readiness checks to ensure traffic will traverse the intended
VPN path. Specifically, it verifies that the tunnel is present and that the namespace’s externally observed identity reflects the
VPN exit (e.g., by querying a public endpoint to learn the egress IP). \tool{} periodically logs the observed exit IP and tunnel status to prevent silent fallback to a non-VPN path.
Only after these checks succeed does \tool{} mark a vantage as ready for measurement execution.

\subsection{Measurements and Trace Capture}
\label{subsec:measurements}
After the Vantage Manager’s checks pass, the Coordinator schedules measurements for the target sites within each namespace. For each website, the Measurement Runner executes \emph{inside the selected namespace}, loads the site, and records an HTTP Archive (HAR) trace. In our deployment, executions are synchronized across namespaces: with $n{=}4$ vantages, we run the Measurement Runner concurrently for each website. The Coordinator can also be configured to introduce a configurable delay between launches, providing time separation if desired. While this not necessary at our current scale, we recommend adding such delays for future studies that use larger $n$ to reduce overloading webservers or triggering rate-limiting effects.

\textbf{Browser-driven measurements: }
The Runner launches a fresh, automated browser instance for each website in each run to make sure no state carry-over (e.g., cookies and caches) across sites
and vantages. It then visits and loads a predefined list of websites under consistent execution settings (timeouts and bounded
retries), enabling controlled comparisons across VPN providers and regions. 

\textbf{HAR recording and metadata logging: }
During each page load, \tool{} records an HTTP Archive (HAR) trace capturing the client-observed request/response activity. 
These traces support downstream infrastructure analysis, including the set of hosts contacted during a page load, the organizations those hosts belong to (via AS-to-organization mapping~\cite{caidaas2org}), and their geographic location (via IP geolocation). Each run also outputs structured metadata, including the timestamp, observed egress IP, and the initial and final websites loaded (after redirects), to support filtering and subsequent analysis.

%% file: re-struc/method.tex
\section{Data Collection}
\label{sec:method}

We created a large-scale dataset by accessing 1,000 websites in each of 14 countries using four different VPN providers. For each country we used \tool{} to connect to four different VPNs, load the websites and recorded HTTP Archive (HAR) files for every website. Each unique website is loaded concurrently across all four VPNs in each country to reduce temporal distortions.
To ensure comprehensive and robust data collection, each website was loaded 25 times from each VPN over a three-week period.

\subsection{VPN Selection and Validation}
\label{VPN-Selection-and-Validation}

We selected four VPN providers, choosing a mix of widely used commercial services and those featured in previous research studies~\cite{kumar2024choices,khan2023stranger,RKumar-SIGMETRICS23}. Specifically, we included NordVPN and Surfshark based on their use in prior academic work, and ExpressVPN and ProtonVPN due to their popularity among users globally.

\textbf{Setup:} Using \tool{}, all VPNs were configured using \texttt{.ovpn} configuration files, with connections established over OpenVPN using UDP. We disabled any built-in ad-blocking or tracking protection features provided by the VPN that could interfere with measurements. We also confirmed with VPN providers that no additional filtering was active during manual configuration.\footnote{For instance, ExpressVPN support clarified that their ad blocker is only available through their VPN apps, not OpenVPN manual setup.}

To minimize variability unrelated to VPN behavior, we deployed all measurements on machines with identical hardware and system configurations. For each VPN, the same country \emph{.ovpn} configuration file was used consistently for three weeks (30 Mar. to 18 Apr., 2025) of measurement.

\textbf{Country Validation:} Previous studies have shown that VPNs sometimes misrepresent their location and may not be hosted in the advertised countries~\cite{weinberg2018catch}. We verified the location of each VPN server using a method similar to prior work~\cite{kumar2024choices,gamero2025localization,weinberg2018catch}.
Each time we connected to a VPN server, we recorded its IP address, along with the IP address before and after loading a website, to detect any changes.
For each VPN-IP, we fetched the geolocation using IPinfo~\cite{ipinfo}. If the reported geolocation matched the advertised country, we then launched a latency probe using RIPE Atlas~\cite{ripeatlas} from a probe located in the same city.
We used a threshold of <1 millisecond for this probe to correspond to a maximum physical distance of approximately 200 km~\cite{katz2006towards}.

We validated each VPN-IP within 48 hours of the original measurement. %
If either the geolocation or latency test failed, we discarded all data collected during the session.
Across all VPN providers, we saw 221 unique IPs. Of these, 24 failed the geolocation check and 24 failed the latency test, leaving 173 valid VPN IPs. Only data collected through these 173 valid IPs was used for further analysis.
These inclusion criteria are intentionally conservative: we prefer to exclude data rather than risk reporting misleading results confounded by misrepresented locations.

\subsection{Country Selection}

Our four VPN providers offered servers in 76 countries. We first filtered these countries based on whether they had been included in previous research studies~\cite{kumar2024choices}, resulting in a pool of 49 eligible countries. From this set, we selected 14 countries, aiming for a geographically balanced mix while also considering practical constraints.
Most commercial VPNs allow at most 10 simultaneous connections/devices, which constrained how many locations we could probe in parallel. We focused on 14 geographically diverse countries rather than a larger set with sparser coverage.

North and South America had only three eligible countries from each continent. To maintain regional coverage, we included all of them.
Europe was more heavily represented, with 23 countries in the eligible pool. We divided the continent into four regions and  randomly selected two countries from each.
This approach resulted in a geographically diverse set of 14 countries for our study, found in Table~\ref{tab:selected_countries}.

\begin{table}[t]
\centering
\begin{tabular}{|l|l|}
\hline
\textbf{Continent} & \textbf{Countries} \\ \hline
North America & United States (US), Mexico (MX),\\
 &  Canada (CA)\\ \hline
South America & Brazil (BR), Argentina (AR),\\
 & Chile (CL)\\ \hline
Northern Europe & Sweden (SE), United Kingdom (UK)\\ \hline
Western Europe & France (FR), Switzerland (CH)\\ \hline
Eastern Europe & Hungary (HU), Czech Republic (CZ)\\ \hline
Southern Europe & Italy (IT), Spain (ES)\\ \hline
\end{tabular}
\caption{Selected countries}
\label{tab:selected_countries}
\end{table}

\subsection{Website Set}

We look at three classes of websites per country (1,000 sites total): 100 country-specific government domains, 100 country-specific regionally popular domains, and a stratified Tranco sample of 800 globally ranked sites (identical across countries).
Combining the global list with government and regional domains balances breadth and relevance across heterogeneous sets, supports generalization, and reduces sampling bias.

\textbf{Government Websites: }
We rely on a previous study~\cite{kumar2024choices} that provides a list of government domains across multiple countries.
For each country, we perform random sampling without replacement to select 100 government websites.

\textbf{Regional Top Websites: } To sample the top websites for a specific country, we aggregate data from four sources: Similarweb~\cite{similarweb}, Semrush~\cite{semrush}, Ahrefs~\cite{ahrefs}, and Cloudflare Radar~\cite{cloudflareRadar}. These sources provide most visited websites within a given region and have been used in previous studies.
We remove previously sampled government websites from this dataset to avoid redundancy, then perform random sampling without replacement to select 100 websites.

\textbf{Tranco Stratified: }
The Tranco list provides a global research-oriented website popularity ranking (1M-site). To ensure a diverse and representative dataset, we stratify ranks into four groups \(S_1\)-\(S_4\) (as in prior work~\cite{lee2022password}): \(S_1\) (1--10K), \(S_2\) (10{,}001--100K), \(S_3\) (100K--500K), and \(S_4\) (500K--1M).
Within each stratum, we perform weighted sampling without replacement with weights inversely proportional to rank, \(w_i=1/r_i\).
This weighting scheme favors more popular websites (i.e., those with lower rank values). %
We select 200 sites per stratum (total \(800\)), favoring more popular sites while maintaining balance across popularity tiers and avoiding over-selection of mid/low-ranked sites.

%% file: re-struc/data.tex
\subsection{Data Gathering and Filtering}
\label{sec:data-filter}
We deployed \tool{} on physical machines provided by \cloudlab{}~\cite{Duplyakin+:ATC19}. Each country was assigned a dedicated machine, and on each machine, four VPN connections to that country were established in separate network namespaces. 
Chrome instances were launched within each namespace to load websites and record data using the corresponding connected VPN. Each instance waits for 30 seconds to allow page loading, approximately twice the typical load time~\cite{butkiewicz2011understanding} and is forcefully terminated after 120 seconds if it becomes unresponsive or fails to progress for any reason.
All machines used were identical in hardware configuration and no anti-bot techniques were employed in the Chrome instances. 

We define \textbf{Source Country} as the country where the VPN is connected. A \textbf{Snapshot} refers to a single measurement instance where the tool collects data from a predefined set of websites using one VPN in one country. A \textbf{Measurement-run} consists of five consecutive snapshots taken at regular intervals within a day, executed in parallel across all VPNs for a source country. 

We conducted five measurement runs over the course of three weeks, with each run spaced approximately three days apart. This was expected to produce a total of 280 measurement runs (14 countries x 5 runs x 4 VPNs). However, 10 runs failed due to unexpected disruptions.\footnote{We did not observe any clear pattern in the failures, except that at least one run failed for each of the four VPNs in the case of Chile. For the other countries, only a single run failed for one VPN.} All VPN sessions and experiments were initiated simultaneously within a given source country to ensure synchronization and eliminate order-related biases.

\textbf{Data Filtering: }
We apply strict filtering to minimize sources of variability unrelated to VPN behaviour, using both network-level signals (VPN IP, request counts) and content-level signals (page titles, redirect targets).

We first filtered all sessions in which the VPN was not properly connected or the IP address changed during the session; this filter removed 19,000 of our 1.228 million snapshots.
We next applied the country validation process described in Section~\ref{VPN-Selection-and-Validation}, which
reduced the dataset size to 606K snapshots. This substantial drop is largely due to entire VPN providers being invalid in specific countries (e.g., ExpressVPN and Surfshark in Mexico). %

We also applied filtering based on the fetched contents.
First, we eliminated snapshots of non-meaningful error pages (like "404 Not Found" or "Access Denied"). We used an LLM (ChatGPT~\cite{openai2023chatgpt}) to analyze 2,700 unique page titles, identifying 121 likely error titles, which we manually verified.
Second, we removed snapshots with zero recorded network requests. Finally, we filtered out cases of website mismatches where the initial and final URLs differed.
After all filtering steps, 510K valid snapshots remained. This comprehensive filtering reduced the total number of unique websites, leaving an average of 556 valid websites per country, ranging from a low of 514 in Argentina to a high of 586 in Sweden.

\textbf{Artifact Availability:} Upon publication, we will release \tool{} along with a sanitized subset of our dataset, including derived data used in the analyses. IP prefixes and other potentially identifying fields will be anonymized and documentation of redactions will accompany the release.

%% file: re-struc/rq1.tex
\section{RQ1: Endpoint IPs and countries}
\label{sec:rq1}

\begin{figure*}[t]
    \centering
    \includegraphics[width=\textwidth]{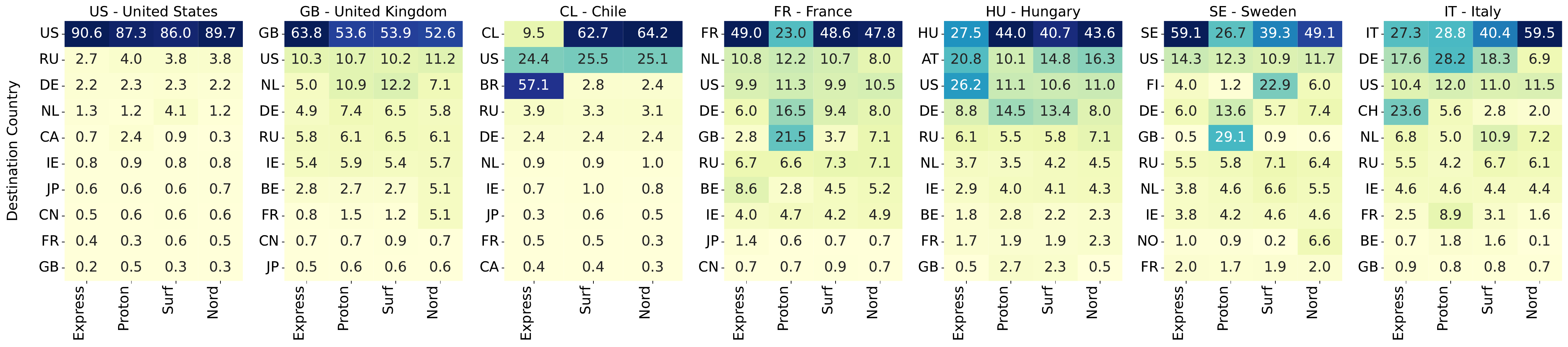}
    \caption{Top-10 destination-countries by VPN, for a representative subset of seven countries.}
    \label{fig:request-distribution}
\end{figure*}

We first ask whether VPN choice within a fixed source country changes \emph{where} a page's traffic lands. We answer this with two complementary measurements that share the same vantages but observe different layers of the path. The first is a controlled \emph{DNS experiment}: from each (VPN, country) combination we issue A-record lookups for a fixed domain set using the resolver each VPN configures by default, and compare the returned IP sets pair-wise across providers. This isolates the answer the vantage receives, before any browser or page-load behavior intervenes. The second is \emph{browser-driven measurement} that \tool{} performs from the same vantages: we collect the IPs that actually served real page loads, geolocate them, and compare destination-country distributions across providers. This shows how the DNS-layer differences manifest in the destination graph a page reaches end-to-end.

The two views answer linked but distinct questions. The DNS experiment tells us whether the vantage \emph{receives} different IPs (\S\ref{sec:dns-experiment}); the browser-driven view tells us whether those differences \emph{propagate} into the country a page contacts in practice (\S\ref{sec:infra}). Throughout RQ1 we keep the two layers separate so that the country-level effect can be read against the DNS-layer effect that produces it.

\subsection{DNS Resolution Experiment}
\label{sec:dns-experiment}

\textbf{Domain samples.}
We examine VPN-induced DNS divergence on two samples, both pulled from the HAR files recorded by \tool{}: a 25K-domain sample drawn from the Tranco top-1M list (\emph{initial domains}) and a 25K-domain sample of \emph{dependency domains} \emph{i.e.}, web dependencies (the JavaScript, analytics, font, and embedded-asset hostnames that real page loads actually pull in).
Running both samples through the same apparatus lets us answer two questions at once: \emph{what} the structural sources of the variability are, and \emph{how} the magnitude of each source depends on whether the observable is an origin endpoint (initial domains) or a CDN-fronted dependency endpoint (dependency domains).

\textbf{A-record comparison.}
For each (VPN, country) combination we query domain samples using the DNS resolver the VPN configures by default; this is the configuration a user or a measurement study would inherit without additional effort. For each domain we then compare the returned A-record sets pair-wise across every pair of VPN providers operating in the same country. A pair is counted as IP-divergent when the two A-record sets are not identical. This lookup gives us the full set of IPs the resolver advertises for each domain, rather than the single IP a browser load happens to contact. Comparing sets rather than single endpoints increases our coverage of the infrastructure each vantage is steered toward.

\textbf{IP-level divergence under each VPN's default DNS.}
Under each provider's default DNS configuration, the aggregate cross-VPN exact-match rate is 88.0\% on the initial sample (range 76.7\% to 94.0\%) and 69\% on the dependency sample (range 59.9\% to 77.4\%).
We also computed Jaccard-based similarity across the same pairs: relaxing the threshold from Jaccard $\geq$ 0.90
down to $\geq$ 0.50
moves aggregate agreement rates by at most half a percent, showing that the disagreements are near-binary: when two VPNs disagree on a domain's IPs, they disagree almost entirely, with disjoint rather than off-by-one IP sets.

This experiment shows that \emph{different VPNs return different IP addresses for the same web pages}. For initial domains roughly one in eight VPN-pair lookups disagrees, and on dependency domains the rate rises to nearly one in three. Whether these DNS-layer disagreements actually move the destination geography a page reaches---or wash out at coarser resolution because the diverging IPs share a country---is what the browser-driven view turns to next.

\subsection{Endpoint Country Distribution}
\label{sec:infra}

We now turn from the DNS experiment to the browser-driven measurement: instead of comparing the IPs each VPN \emph{receives} from a resolver, we compare the IPs that actually \emph{served} real page loads under \tool{}, geolocate them, and ask whether the destination-country distribution differs across VPNs within each source country.

We perform IP geolocation using IPinfo~\cite{ipinfo} for each IP, and compute the percentage of traffic from each VPN (per source country) that is routed to server endpoints located in different countries.\footnote{Prior work finds that commercial IP geolocation is generally reliable at the country level, with high cross-database agreement on IP-to-country mappings~\cite{huffaker2011geocompare,callejo2022deep}. This level of accuracy is sufficient for our analysis, which focuses on country-level endpoint provenance rather than fine-grained city-level location.}
To reduce variability introduced by anycast IPs, we leverage recent data from the Manycast2 project~\cite{sommese2020manycast2} to identify and exclude anycast IPs in our dataset from this part of the analysis. In total, we observed 11.15 million unique URLs and 149K unique IPs, of which 3561 were identified as anycast.

We use a heatmap (Figure~\ref{fig:request-distribution}) to analyze the top 10 destination countries for network requests from 7 of our countries. Each section highlights the percentage of total requests routed to these destinations, showing a clear comparison of how VPNs may affect the measurements by centralizing or diversifying endpoint locations.
The United States and United Kingdom show consistency across all VPNs, with most requests served within the source country.

Other countries reveal substantial \textit{VPN-dependent} behaviour. For instance, in Chile, while Nord and Surf routed over 60\% of traffic to Chile itself (suggesting local endpoints or effective regional peering), ExpressVPN routed 57.1\% to Brazil. This difference underscores VPN-specific routing strategies.\footnote{This may be due to differences in infrastructure, agreements with CDNs, or regional exit node availability.}
Traffic from France shows that Proton routes requests across a broader mix of countries compared to other VPNs.
Similar trends of VPN affecting the destination country of web traffic can be seen in others such as Express in Hungary, Proton in Sweden and Nord in Italy.

Across the 14 countries we studied (Appendix~\ref{app:remaining-countries} shows the remaining 7), we found six where VPN choice has minimal impact, and eight with significant differences.
These results indicate that different VPNs connect to different global infrastructures even when originating from the same source country. Consequently, measurement outcomes (e.g., latency, CDN reach, censorship checks) may vary not because of the website, but due to VPN-induced bias in destination selection.

We further quantify inter-VPN server endpoint differences using Jensen-Shannon Distance (JSD), a symmetric bounded (0-1) divergence,
giving us a statistical test whether
VPN choice meaningfully shifts the destination-country distribution.
Across 130 VPN-pair comparisons in 14 countries, JSDs are modest on average (mean 0.17, median 0.13) but unevenly distributed. Argentina, Chile, Sweden, and Italy account for nearly all of the right tail (means 0.23--0.40), while Spain, Mexico, and the United States stay below 0.10 across every pair. Chile alone holds both extremes \emph{i.e.}, Express vs. Nord at 0.51 and Nord vs. Surf at 0.03, showing that within a single country, the divergence is provider-specific rather than country-wide.
Figure~\ref{fig:jsd_heatmaps} in Appendix~\ref{app:remaining-countries} visualizes JSD and these patterns across all countries.

\textbf{Takeaway.} The two experiments give a consistent picture across layers. The DNS experiment shows that VPN choice shifts the IPs returned for a domain---$\sim$12\% pair-wise divergence on initial domains, $\sim$31\% on dependency domains, with disagreements that are near-binary rather than off-by-one. The browser-driven view shows that this DNS-layer divergence partly propagates into the destination country a page contacts, but unevenly.
Where the propagation does occur, it reshapes the destination graph enough to alter downstream inferences about cross-border exposure and data localization, implying that previous studies using VPN-based vantage points to reason about national data-flow patterns may have reached conclusions that are conditional on the VPN provider rather than properties of the country itself.
Whether this same divergence exists at the organizational level is the question we take up next.

%% file: re-struc/rq2.tex
\section{RQ2: Hosting organization}
\label{sec:rq2}

\begin{figure*}[t]
    \centering
    \includegraphics[width=0.9\textwidth]{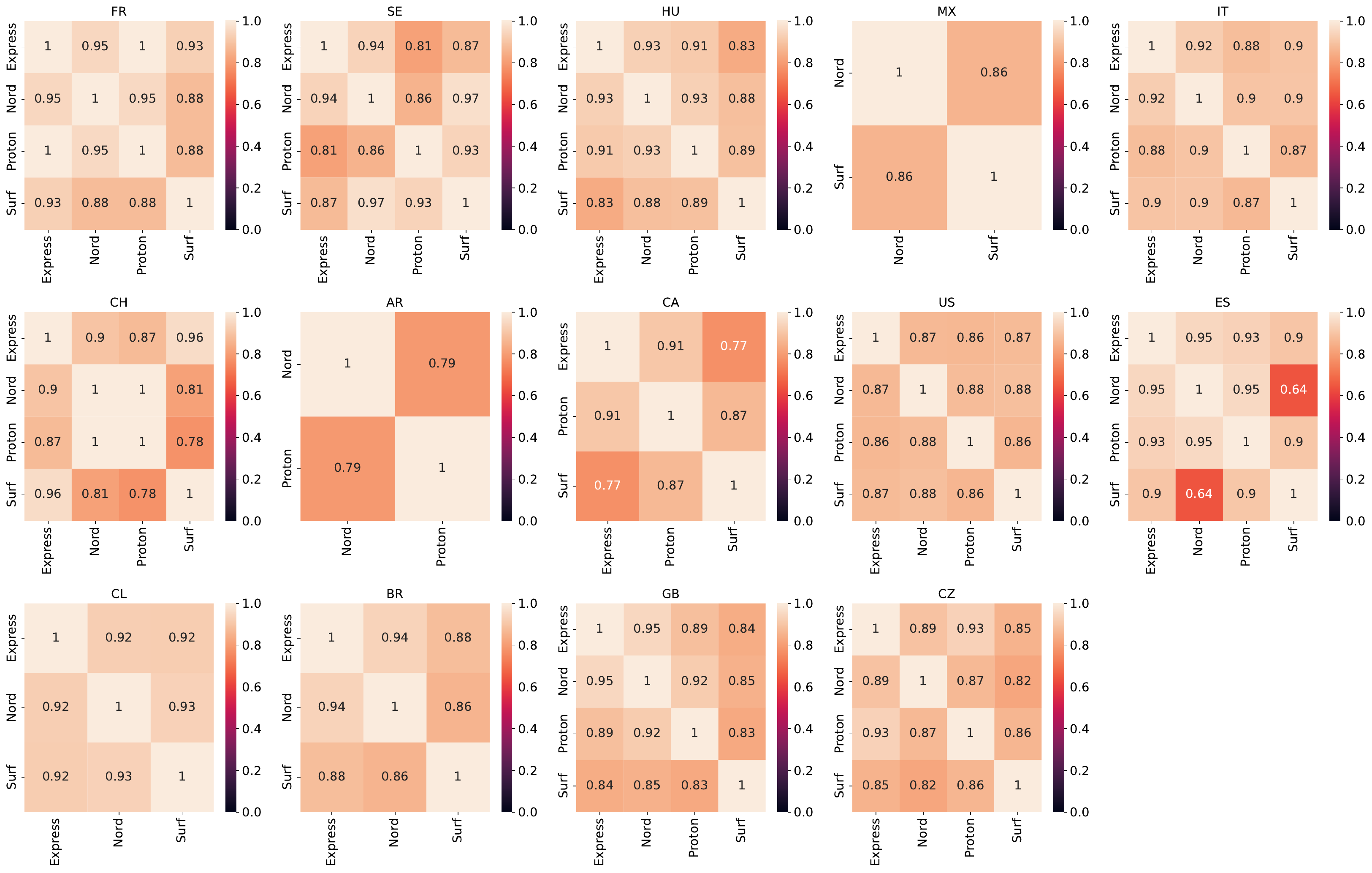}
    \caption{Pair-wise Jaccard similarity of the set of hosting organizations reached across VPNs, averaged across all websites in each source country. 
    }
    \label{fig:vpnorg_js_heatmaps_per_country}
\end{figure*}

RQ1 showed that VPN choice shifts both the IPs returned for a domain and the country those IPs geolocate to. We now ask whether the shift persists when we consider the \emph{organization} that owns each IP. For example, the same domain may resolve to different IPs across VPNs (IP disagreement) while all of these IPs are hosted by Cloudflare (organization agreement); whether VPN choice ``matters'' for an inference therefore depends on the resolution at which the inference is drawn.
We answer this with the same measurements used in RQ1, re-projected onto the organization layer.

\subsection{Hosting Provider}
\label{sec:hosting-org}

We start with the browser-driven view: for each (website, country) combination we take the set of hosting organizations \tool{} actually contacted under each VPN and compare them pair-wise.

Figure~\ref{fig:vpnorg_js_heatmaps_per_country} shows pair-wise Jaccard Similarity of the hosting-organization sets reached across VPNs, averaged across all websites in each source country. \emph{VPN selection rarely changes which organization serves a site}. Cross-VPN agreement is high overall: mean JS is 0.88 across our 14-country dataset, and 43\% of websites see perfect overlap (JS = 1) across all VPN pairs. Most countries show a high proportion of VPN-pair combinations with agreement over 0.75. For example, Sweden and Hungary have over 89\% of their measurements above this threshold, suggesting strong consistency in hosting organizations across VPNs for most sites. Conversely, Mexico and Canada exhibit lower alignment, with only 76--77\% of JS values above 0.75, indicating more variation in observed behavior. The percentage of low-similarity instances (JS < 0.5) remains low across all countries (generally below 2.5\%) showing that extreme inconsistencies are rare but do occur.

However, specific cases reveal notable deviations. In five countries, the lowest JS values were exactly zero, meaning the VPN pairs accessed the same website through \emph{completely disjoint} sets of hosting providers. For example, in France and Sweden, we observed a JS of 0.0 between Surf and Nord when accessing netflix.com and skatteverket.se, respectively. Such divergence implies that VPNs may route traffic through \emph{different regions or CDNs}, leading to distinct infrastructure being used even within the same country for the same site.

The page-level view aggregates over the full set of organizations a real page load contacts---origin and dependencies together---and tells us whether that aggregate set agrees across VPNs. It does not, however, separate the two surfaces. To see whether dependency endpoints behave differently from origin endpoints at the organization layer, we turn to the DNS experiment.

\subsection{DNS Experiment: Organization View}
\label{sec:org-dns-results}

We now revisit the DNS experiment from \S\ref{sec:dns-experiment}, this time mapping each returned IP to its parent organization and counting a VPN pair as agreeing if their organization sets overlap. Reporting separately on the initial and dependency samples lets us see whether the dependency-driven amplification observed at the IP level is preserved or absorbed at the organization layer.

Here, the picture changes sharply relative to the IP-level view in RQ1---\emph{initial and dependency domains both see significant agreement}.
At the IP level, the dependency sample diverges roughly $2.6\times$ more than the initial sample: 12\% of VPN-pair lookups disagree on initial domains, rising to 31\% on dependency domains. At the organization layer, this amplification almost completely vanishes. Agreement is 98\% on the initial sample and 96\% on the dependency sample---equivalently, organization-level \emph{disagreement} rises from 2\% to 4\%, a $2\times$ increase. 
VPN-induced DNS divergence is therefore overwhelmingly an intra-CDN replica-selection phenomenon rather than a cross-provider routing one: different VPN exits are steered to different Cloudflare, Amazon, or Google edges, not to different hosting companies.

\subsection{Takeaway}
These two experiments give a consistent picture at the organization layer. The browser-driven view reports a mean JS of 0.88 across all 14 countries, with 43\% of websites seeing perfect overlap across every VPN pair; the DNS view reports 98\% organization-level agreement on initial domains and 96\% on dependency domains.
This suggests that prior work on third-party dependency and centralization is likely to remain qualitatively valid under VPN-vantage variability, since the set of organizations a country's web traffic depends on is largely robust to VPN choice.
RQ3 (\S\ref{sec:rq3}) examines which layers of the VPN-to-server path produce these IP- and country-level disagreements while leaving the organization layer largely intact.

%% file: re-struc/rq3.tex
\section{RQ3: Layers Driving the Variability}
\label{sec:rq3}

Sections~\ref{sec:rq1} and~\ref{sec:rq2} establish that VPN selection measurably shifts which endpoints a page reaches and that the magnitude of the shift depends on the resolution of the observable.
In this section we ask where that variability comes from, and for each factor we note how much of it is actually under a measurement study's control.

A VPN session is a stack of decisions. The client connects to a server; the server sits in some physical location; the tunnel emerges into some hosting network with some ASN; DNS queries are answered by some resolver; that resolver's answer is ultimately produced by some authoritative nameserver making a steering decision; and the resulting packets are forwarded along some BGP path to one of possibly many replicas of the target service. Any of these layers could drive the divergence we observe, and the layers are not independent: a change at one level (say, which ASN the VPN egresses through) propagates downstream into all the others --- which IPs a CDN returns, which PoP receives the packets, and therefore which country and which organization a domain associates to.

We distill this stack into six candidate factors, grouped into three layers:

\begin{itemize}[leftmargin=*, noitemsep, topsep=2pt]
  \item \textbf{Vantage identity} --- where the VPN exit actually sits in the physical and network world:
  \textbf{C1} virtual exit locations (exits not physically in the advertised country);
  \textbf{C2} geographic dispersion of same-country exits (exits in different cities).

  \item \textbf{Name resolution} --- how domain names are translated to IP addresses:
  \textbf{C3} DNS resolver differences (each provider ships a different default resolver);
  \textbf{C4} DNS interception (provider transparently redirects port-53 traffic).

  \item \textbf{CDN routing} --- how CDNs steer requests to replicas based on the forwarding topology:
  \textbf{C5} CDN steering keyed on the exit ASN and IP prefix;
  \textbf{C6} peering topology and BGP routing to anycast replicas.
\end{itemize}

C1, C2, and C4 have been touched on in prior work on the commercial VPN ecosystem~\cite{weinberg2018catch,khan2018empirical,ramesh2022vpnalyzer,pearce2017global}; we revisit each in our setting and highlight which factors are under study-author control and which are structural properties of the VPN-to-endpoint path.

\subsection{Vantage Identity (C1, C2)}
\label{sec:vantage-location}

The vantage-identity layer, where the VPN exit actually sits in both country (C1) and city (C2), is well-studied in the VPN-measurement literature~\cite{weinberg2018catch,khan2018empirical,ramesh2022vpnalyzer}.

\textbf{C1: Virtual exit locations: }
Commercial VPN providers sometimes advertise exits in countries where they have no physical presence, using geolocation-database entries to make a server elsewhere appear to be in the advertised country. We verified every exit against IPinfo and cross-checked with latency checks (\S\ref{sec:method}), in the browser experiment, 48 of 221 exits (21.7\%), and in the DNS experiments 15 of 118 exits (12.7\%) were outside the claimed country.

\textbf{C2: Geographic dispersion of same-country exits: }
    Even with verified exits, two servers advertised for the same country may sit in different cities, and CDN steering to nearby replicas will produce legitimate and provider-extrinsic divergence. Of our 14 countries, 12 have all VPN exits in the same city; the two exceptions span $\sim$3{,}200\,km and $\sim$4{,}400\,km between their furthest exits. When we performed DNS lookups for our domain samples across these exits, geography contributed less than one might expect. In the first multi-city country, same-city pairs diverge at 9.0\% while cross-city pairs (across $\sim$3{,}200\,km) diverge at 9.8--10.2\%, a slim gap despite the separation. The second multi-city country shows a larger effect: a $\sim$540\,km pair diverges at 17.5\% while pairs spanning the full $\sim$4{,}400\,km diverge at 23.4--23.5\%, a six-point geographic component.

\textbf{Implication: }
A misplaced exit produces endpoint-country distributions that appear to violate locality for reasons unrelated to the  sites being measured, and standard geolocation verification is sufficient to eliminate the confounding factor. Both C1 (virtual exits) and C2 (city dispersion) are controllable by a measurement study via geolocation verification and same-city reporting, and neither can explain the persistent same-city divergence. We hold both constant in the rest of this section.

\subsection{Name Resolution (C3, C4)}

\label{sec:name-res}
Each VPN provider has its own default DNS configuration, and the configurations differ architecturally across providers: some run a dedicated resolver reachable from the tunnel, others run an internal resolver on the tunnel interface itself, and some transparently redirect all port-53 traffic to their own resolver regardless of what the client configures~\cite{khan2018empirical,ramesh2022vpnalyzer}. These implementation choices can produce divergent A records for the same domain across providers even when nothing else about the vantage differs.

The baseline DNS-divergence numbers (88\%/69\% IP-level and 98\%/96\% organization-level on initial/dependency samples) were reported in \S~\ref{sec:rq2}. To understand \emph{why} the divergence exists, we traced queries in more detail.

\textbf{Tracing where DNS queries actually travel.}
    To record DNS resolution paths,
    we registered DNS canary tokens~\cite{thinkst-canarytokens}: uniquely-named domains served by authoritative nameservers under our control. Every query to these domains is logged along with the source IP of the recursive resolver that reached us and the specific sub-label that was queried.
Because each (VPN, country) combination uses a distinct sub-label, every hit can be attributed to the exact experimental configuration that produced it.
From each (VPN, country) combination we issued a query for one of these canary domains, naming a unicast resolver located outside the country as the resolver. If the provider honoured the request, the canary would see the query arrive from that unicast resolver; if the provider intercepted, the canary would see whatever resolver the VPN actually used instead.

\textbf{Different VPNs, different DNS paths: }
    The picture is not uniform. Across 14 countries and 4 providers, only ProtonVPN consistently honors the resolver the client names: every one of its canary hits arrives from the unicast resolver we specified, in every country tested. The other three providers (ExpressVPN, NordVPN, and Surfshark) transparently intercept port-53 traffic regardless of what the client configures, and the interception mechanism itself varies by country. In Italy, Spain, Czechia, Canada, and France, the canary receives queries from IPs inside the VPN exit's own /24 prefix. Surfshark's exit IP is itself the resolver, indicating the provider runs its own recursive resolver co-located with the exit. In Brazil, the UK, the US, Sweden, Hungary, Argentina, Chile, and Mexico, the canary instead records queries from Cloudflare anycast (AS13335), indicating the provider intercepts locally and forwards upstream to 1.1.1.1.
Interestingly, within a given country the three intercepting providers converge on the same interception mechanism, a pattern that may suggest in-country DNS infrastructure being leased or provisioned at the country level rather than decided independently by each provider.

\textbf{Implication: }
For three of the four providers we study, DNS resolver selection is not a controllable variable at the client: the resolver is replaced by whatever the VPN decides to use. The baseline divergence we observe (\S\ref{sec:rq2}) is therefore caused by each provider's own DNS infrastructure, and cannot be narrowed by pointing the client at a different resolver. The 98\%/96\% organization-level agreement softens how much this matters in practice: different VPN providers end up at different Cloudflare or Amazon edges, but not at different companies.
Studies asking which \emph{organization} serves a request can treat the name-resolution layer as largely VPN-robust. 
Studies that depend on \textit{which IP --- and in which country}--- serves a request need to 
(a) identify and include a non-intercepting provider as a reference point, or (b) treat the DNS answer as produced by the VPN operator's in-country resolver infrastructure rather than by the client's resolver configuration.

\subsection{Replica Selection (C5, C6)}
\label{sec:replica}
\label{sec:replica-probe}
\label{sec:c5}
\label{sec:c6}

CDNs steer requests to replicas based on the IP address they see from the client (in this case, the one assigned by the VPN provider). Our final two sources of variability look the granularity of this steering and its effects on replica selection.

\textbf{C5: CDN Steering on Exit ASN and Prefix}
A CDN's authoritative nameserver returns an IP chosen based on the querier's apparent network location, inferred from the recursive resolver's source IP or, when present, from the EDNS Client Subnet (ECS)~\cite{rfc7871} prefix forwarded on the client's behalf. Different VPN providers egress through different hosting companies, so the source IP and ECS prefix seen by a CDN differ per provider even when the exit city is the same.

\textbf{Same city, same AS, still divergent:}
To see how fine-grained this steering is, we decompose the baseline pair-wise divergence rate by whether both VPNs in a pair egress from the same city and the same hosting AS. Across our 14 countries, 15 VPN-pair configurations share the same city and the same hosting AS --- e.g., UK Nord--Surf on AS25369 (Hydra Communications), AR Express--Nord on AS396356 (Latitude.sh), ES Express--Nord on AS212238 (Datacamp), IT Express--Surf on AS9009 (M247), FR Nord--Surf on AS25369. 
Sharing an AS is not enough to eliminate the disagreement: these five example pairs show IP-level DNS divergence rates of 6.3\%, 6.7\%, 8.8\%, 9.6\%, and 15.9\% respectively --- i.e., the fraction of domains for which the two VPNs' A-record sets are not identical, using the same metric as in \S~\ref{sec:dns-experiment}.
Of the 15 same-AS pairs, only 2 additionally share the same /24 prefix; on those two the divergence collapses toward zero, while the other 13 (same city, same AS, different /24) show the 6--16\% residual above. This is the cleanest indicator we have that CDN steering operates at the prefix granularity: within a single AS, VPN providers typically occupy different /24 prefixes, and CDN steering databases are prefix-keyed rather than AS-keyed.

\textbf{C6: Peering Topology to Anycast Replicas}
Even when two providers resolve a domain to the \emph{same} IP, they may not reach the same physical server. Large CDNs announce the same anycast prefix from hundreds of PoPs, and BGP routing from each upstream network selects a different one~\cite{cicalese2015characterizing,sommese2020manycast2}. DNS convergence is necessary but not sufficient for routing convergence: providers that receive identical A records can still land on PoPs hundreds of kilometers apart, invisible to any analysis that stops at the DNS answer.

To identify which physical PoP each vantage actually reaches, we run a targeted HTTPS probe. %
    From the baseline DNS results on both samples (initial and dependency URLs) we extract the set of IPs that (a) were returned for the same domain by at least two verified providers and (b) fall within a known anycast prefix according to the Manycast2 database~\cite{sommese2020manycast2} (5K IPv4 prefixes). We restrict the analysis to three major CDNs (Cloudflare, CloudFront, and Fastly) because their serving edges return a self-identifying PoP header that we can attribute with certainty to a specific physical facility\footnote{Other CDNs either do not expose such a header or expose one that is not reliably decodable.}
Across all 14 countries, 65\% of probed anycast IPs were served by one of the three CDNs; the analyses in the remainder of this section are on that subset.

We classify each probed IP as \emph{Exact Same} (all providers return identical identifiers), \emph{Same City} (identifiers share the IATA airport-code prefix but differ in the server-instance suffix)
\emph{Different} (IATA prefixes differ),
or \emph{Can't Decide} (fewer than two providers returned a parseable PoP header).\footnote{We avoid RTT-based classification: tunnel overhead (50--60 ms) dominates the target-to-replica latency difference (5--15 ms).}

\textbf{Three country groups:}
For each country, we probed every qualifying anycast IP with its corresponding SNI (Server Name Indication) on both samples, 65\% were served by one of the three CDNs, and the results below are on that subset.
Three country groups emerge. \emph{Consistently divergent} --- the US, Mexico, Canada, Chile, Czechia, Italy --- route a majority of probed anycast IPs to different cities on both samples. Italy shows 56\% Different on initial and 64\% on dependency, with Milan (MXP) vs.\ Frankfurt (FRA) accounting for most of it --- a 700\,km gap despite both exits being in Milan; the US shows 62\% initial and 70\% dependency. \emph{Consistently convergent} --- Brazil, Switzerland, Spain --- route almost every probed IP to the same replica (BR: 4\% Different on initial, 0\% on dependency; CH: 29\% / 0\%; ES: 16\% / under 1\%). \emph{Mixed} --- UK, France, Sweden, Hungary, Argentina --- show notable divergence on initial but near-zero on dependency (UK 26\% $\rightarrow$ 0\%, France 58\% $\rightarrow$ 0\%, Sweden 44\% $\rightarrow$ 0\%). Can't-Decide rates sit at 22--40\% on initial and 24--38\% on dependency, dominated by targets whose CDN does not expose a parseable PoP identifier.

\textbf{Implication: }
C5 (prefix-level CDN steering) and C6 (peering to anycast replicas) sit below the layers a measurement study can control. They operate on the exit ASN, the exit /24 prefix, and the upstream BGP path --- all properties of the VPN operator's network rather than of the study's configuration.
Unlike C3/C4, C6 in particular is \emph{not} attenuated at the organization level: the PoPs on either side of a divergence are owned by the same CDN company but may sit in different jurisdictions, with different transit paths and different middleboxes, so a study that treats DNS-level organization agreement as evidence of jurisdictional agreement will understate cross-border exposure.

%% file: discussion.tex
\section{Implications for Past VPN-Based Studies}
\label{sec:disc-past}

Our results do not invalidate prior infrastructure-focused findings that relied on VPN vantage points,they identify where the uncertainty in it is concentrated and which claims are affected.
The portion of each finding that lives on the dependency domains is more VPN-dependent than has been acknowledged. The portion that lives on the initial domain is relatively robust and the portion that depends on the IP or country level of an observable is noisier than the portion that depends on the organization level. A second thread runs through any prior work that assumed the client's DNS resolver is a controllable variable: our canary evidence (Section~\ref{sec:name-res}) shows that three of four commercial providers transparently intercept port-53 traffic, so the ``resolver'' such studies thought they were pinning was in many cases the VPN operator's own in-country infrastructure.

\textbf{Third-party infrastructure dependency and centralization studies}~\cite{RKumar-SIGMETRICS23} ask which infrastructure providers a page's resources are ultimately served by. Their core observables live on the dependency domain, where our baseline DNS measurements show a cross-VPN divergence of $\sim$31\% at the IP level and up to 70\% at the peering level in the consistently divergent group (US, IT, MX, CA, CL, CZ). The qualitative finding that the modern web is heavily dependent on a small number of infrastructure providers survives at the organization level, because the 96\% same-organization agreement on dependencies means the \emph{set} of organizations reached is largely VPN-robust. But per-country point estimates of the \emph{degree} of concentration, especially when measured at the IP or PoP level, should be read as conditional on the VPN vantage used, not as a property of the country.

\textbf{Government-hosting studies}~\cite{kumar2024choices} are largely origin-side: who hosts the landing page of a government site is determined by that government's infrastructure choices, and VPN sensitivity is bounded by the initial-domain floor. We would therefore expect the main findings of this line of work to be comparatively VPN-robust, consistent with our 88\% IP-level and 98\% organization-level agreement on the initial domain sample. Where these studies extend to embedded assets or dependency endpoints, the dependency-domain caveats apply, and the jurisdictional framing (``hosted domestically vs.\ abroad'') in particular can shift at the IP or country resolution even when the organizational resolution does not. VPN selection also impacts the interpretations of \emph{data localization}—the extent to which a country’s web traffic remains within national borders. Depending on the provider, one might infer either strong domestic infrastructure or poor localization for the same country and sites. These discrepancies underscore how VPNs can distort visibility into national data-flow patterns, making it crucial to account for VPN-induced variability when using VPN vantage points to reason about localization-sensitive questions, cross-border exposure, or infrastructure dependence--questions that increasingly carry policy and regulatory weight under frameworks such as the GDPR~\cite{EU2016GDPR}.

\textbf{Censorship- and geoblocking-measurement platforms}
use VPN vantage points partly as a way of reaching endpoints~\cite{niaki2020iclab,hoang2019measuring,mcdonald2018403}. Our data does not speak directly to blocking, but the mechanisms we document (C6) may also shape which middleboxes a packet traverses and which IP a blocklist targets. If so, a site reachable from one VPN path and unreachable from another would not necessarily indicate censorship, and distinguishing the two would benefit from multi-provider replication.

\textbf{DNS-measurement studies that assume resolver control} face a second, narrower issue. Our canary evidence shows that three of the four commercial providers we tested transparently intercept port-53 traffic and forward it to their own in-country infrastructure or to Cloudflare, regardless of what the client configures. Any study using commercial VPN vantages to measure resolver-specific behaviour, DoH/DoT performance across regions, upstream TTL handling, or DNS-manipulation mapping needs to verify per (provider, country) that queries actually reach the intended resolver, otherwise the observation reflects the VPN operator's own resolver, not the one under study. Only one of the four providers in our sample (ProtonVPN) honoured client-configured resolvers consistently.

\section{For Future VPN-Based Studies}
\label{sec:disc-future}

The structural nature of C5 (prefix-level CDN steering) and C6 (peering to anycast replicas) means they are hard to engineer away: any study that compares VPN vantages is comparing different network identities, and the CDN layer will treat those identities as distinct. 
The name-resolution layer (C3, C4) is also difficult to control reliably at the client for commercial providers whose default DNS paths diverge from the client's stated preference, as we observed for three of the four providers in our sample. 
The \textit{right response} is to make the variability visible and to interpret it correctly. We distill the following practices from our analysis.

\textbf{Treat VPN choice as an experimental variable: }
Prior work has generally treated VPN selection as an implementation detail, a means of reaching a country. Our results show that the choice itself shapes what is observed. VPN configuration (provider, protocol, version, any on-by-default filtering) should be recorded and published alongside the measurement, the same way a paper would report a browser version.

\textbf{Report vantage identity, not just country: }
Country alone is not a sufficient description of a VPN vantage. Studies should report, at minimum, the exit ASN, the exit /24 (or at least the hosting organization), whether the exit passed independent geolocation verification, and whether the provider intercepts DNS. Every one of these fields was shown in \S\ref{sec:rq3} to have a measurable downstream effect on RQ1 observables, and the DNS-interception status in particular determines whether client-side resolver choices propagate past the tunnel edge.

\textbf{Validate exit locations before using them: }
Our geolocation check excluded 63 of 339 exits (18.6\%) as not physically in the advertised country (48 of 221 in the browser experiment~\S\ref{VPN-Selection-and-Validation} and 15 of 118 in the DNS experiments~\S\ref{sec:rq3}). Future studies should incorporate the same two-step validation we use, IP-geolocation cross-checked against latency from a nearby independent probe and discard exits that fail either test, since measurements from a mis-located exit are simply attributed to the wrong country.

\textbf{Verify resolver reachability when DNS control is required: }
If a study's design assumes the client's configured resolver is the one actually answering queries, that assumption needs to be verified per (VPN, country) pair rather than inherited from the client configuration. Lightweight canary-based probes (Section~\ref{sec:name-res}) can establish whether a given provider honours the client's resolver choice or silently redirects. In our sample only ProtonVPN did so consistently, ExpressVPN, NordVPN, and Surfshark all intercepted, with mechanisms that varied by country.

\textbf{Report which domain surface you measured:} 
The variability we observe is sharply different on initial domains versus dependencies, and which surface a study targets determines how much VPN-induced uncertainty its numbers carry. Measurements of the landing page itself --- its initial endpoint and the organization that serves it --- carry the lower variability we report for the initial-domain sample (88\% IP-level and 98\% organization-level cross-VPN agreement, \S\ref{sec:rq1}, \S\ref{sec:rq2}). Measurements of the embedded resources a page loads --- third-party scripts, fonts, analytics, ad endpoints, CDN-fronted assets --- carry the substantially higher variability of the dependency-domain sample (69\% IP-level agreement, dropping to as low as 30\% Same-PoP at the peering layer in the consistently divergent country group, \S\ref{sec:rq1}, \S\ref{sec:rq3}).

\textbf{Report at the resolution the question demands, and no finer: }
Our IP-ownership analysis shows that organization-level observables are substantially more VPN-robust than IP-level or country-level ones, especially on the initial domain (98\% same-organization agreement against 88\% same-IP). If the research question is really about ``which infrastructure providers serve this country's web traffic'', reporting at the organization level reduces the VPN-induced noise that a raw IP- or country-level report would inherit. 
Where the question is genuinely about country-level jurisdiction, report it at country level but acknowledge the variability across providers, because that is the resolution at which intra-CDN replica selection and peering to foreign PoPs are most visible.

\textbf{Replicate across providers and time: }
A single-provider VPN measurement should be read as one realization of a vantage-conditioned random variable, not as ground truth for the country. Multi-provider replication, with results reported as a distribution rather than a point estimate, turns VPN-induced variability from an invisible bias into an auditable uncertainty band. The 130 pairwise Jensen--Shannon distances we report in \S\ref{sec:infra} are an example of this kind of reporting, the same-AS pair-wise decomposition in \S\ref{sec:c5} is another. Repeating the same measurement across time windows provides the complementary temporal axis, since provider infrastructure, peering, and exit-IP assignments change. 
Studies for which neither kind of replication is feasible can still note that their numbers are conditional on the VPN vantage.

In aggregate these practices do not restore the VPN to the status of a neutral measurement instrument, but they allow the community to quantify and report the uncertainty that the instrument introduces for infrastructure-level inferences about the Web, and in doing so to recover the interpretive value of the measurements it produces.

\section{Limitations}

Our automated browsing relies on Selenium, which can affect how sites are served relative to a real user session~\cite{cassel2022omnicrawl}. Measurements were limited to landing pages, potentially missing infrastructure endpoints that emerge only with deeper interaction~\cite{aqeel2020landing}. To keep cross-VPN behavior comparable, we avoided anti-bot evasion, so some sites may have blocked access or served modified content. We collected data over three weeks across 14 countries, yielding 1.228 million page-load snapshots and 510K valid snapshots after strict filtering. The resulting dataset is sufficient to support statistically significant comparisons in our analyses. 

Our findings are conditioned on several design choices. We study four major commercial providers over OpenVPN and UDP, other providers, WireGuard-based services, and DoH/DoT configurations may behave differently. Country selection is shaped by commercial-VPN server availability, which under-represents regions with sparser VPN footprints and the findings may not extrapolate to countries with materially different peering ecosystems. 
The replica-selection probe is restricted to Cloudflare, CloudFront, and Fastly, the three major CDNs that expose a self-identifying PoP header and other CDNs without such a header are excluded. 
As a first systematic study of VPN-induced effects in web-infrastructure measurement, we believe these design choices are reasonable, transparent, and provide a sound foundation for future, longer-running studies.

%% file: conclusion.tex
\section{Conclusion}

Commercial VPNs are widely used as geographically distributed vantage points for Web measurement, under the assumption that VPNs within a country are interchangeable. 
Our results show that this assumption does not hold. Across fourteen countries and four major providers, VPN choice measurably shifts which endpoints a page load contacts, which countries those endpoints sit in, and which physical replicas serve them, with the effect concentrated on the CDN-fronted dependency domain that dominates modern page loads.
The variability comes from layers below the client: in-country DNS infrastructure that most providers operate regardless of the resolver a client configures, CDN steering keyed on the exit network, and peering paths that route identical DNS answers to different physical replicas. These layers are not under a measurement study's control. 
The effect is also resolution-dependent: inferences at the organization level are largely VPN-robust, while inferences at the IP and country level are substantially more sensitive.
These findings do not invalidate prior VPN-based work, they make visible the uncertainty that has always been present in it. What the community needs is a reporting discipline that treats VPN choice as an experimental variable rather than an implementation detail.

%% file: re-struc/appendix.tex
\section{Appendix}

\section*{Ethics}
Our study is an observational web measurement study and does not involve human subjects. We only access publicly available web pages and to minimize potential impact on websites and network infrastructure, we (i) rate-limit crawls, (ii) distribute requests over time, and (iii) revisit each site only a small number of times per day. We also avoid any active security testing  and restrict collection to the network and browser artifacts needed for our analyses.

\section*{AI Disclosure}
The authors used generative AI tool (Claude Opus 4.7) to revise the text, improve flow, and correct typos, grammatical errors, and awkward phrasing throughout the paper.
Generative AI was also used as part of our methodology, as described in Section~\ref{sec:data-filter}.

\onecolumn
\subsection{Top 10 Destination Countries for web traffic}
\label{app:remaining-countries}

\begin{figure*}[h]
    \centering
    \includegraphics[width=\textwidth]{figures/jsd_heatmap_row_of_7.pdf}

    \vspace{0.5cm} %

    \includegraphics[width=\textwidth]{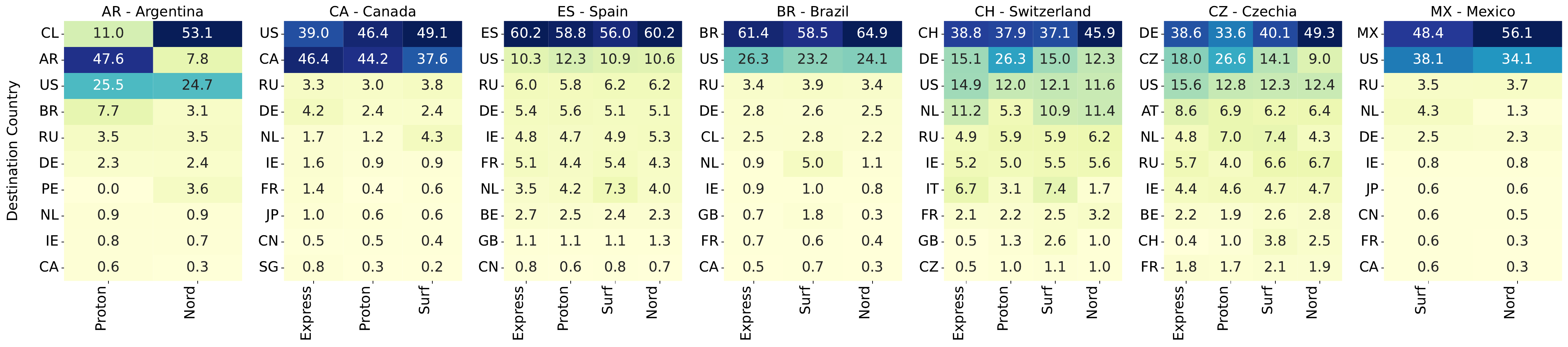}
    \caption{Top-10 destination-country distribution of contacted endpoints by VPN, for all fourteen source countries. Rows are destination countries and columns are VPN providers.}
    \label{fig:request-distribution-all}
\end{figure*}

\subsection{Jensen--Shannon Distance (JSD) of Infrastructural Endpoints}
\label{app:jsd-infra}
\begin{figure*}[h]
    \centering
    \includegraphics[width=0.8\linewidth]{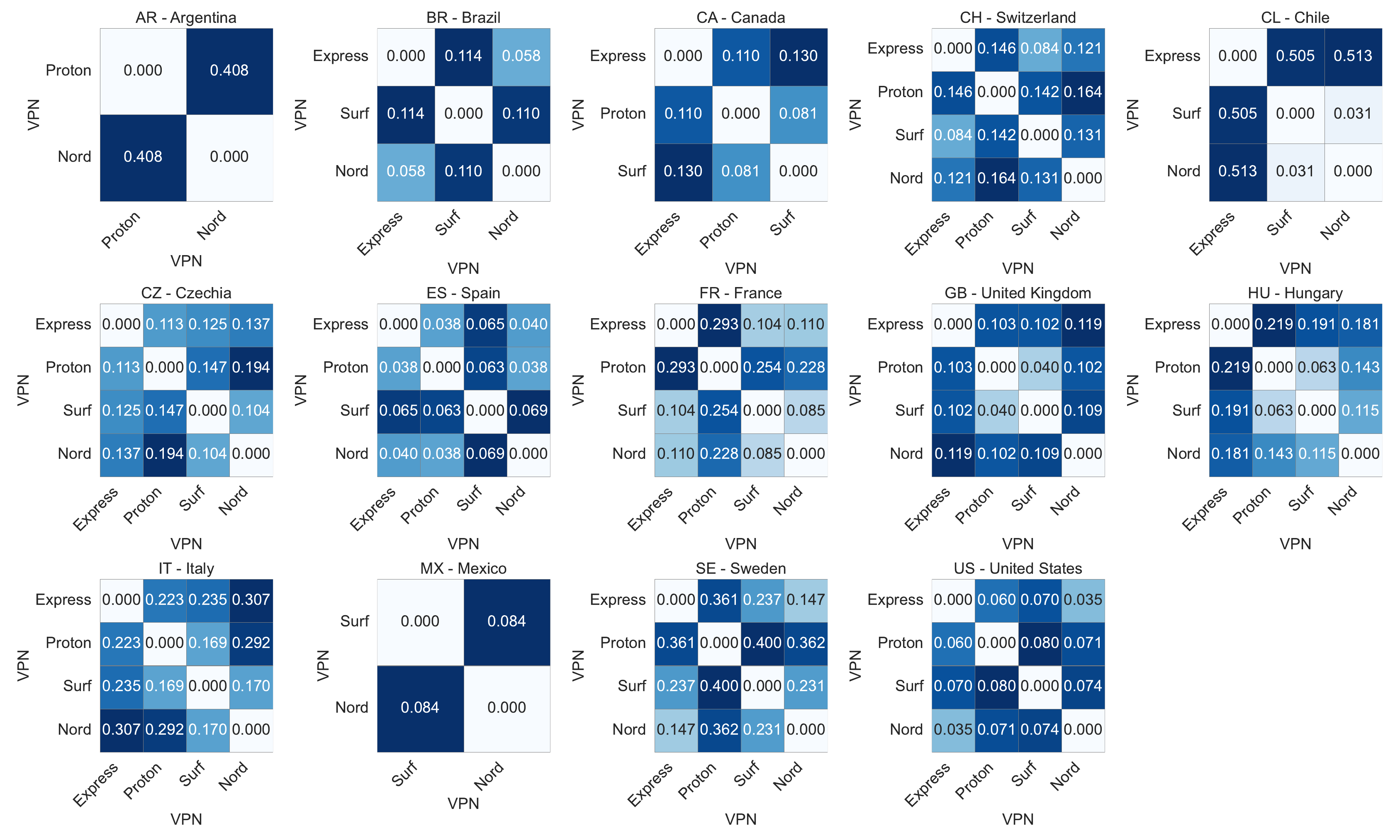}
    \caption{
        Pair-wise Jensen--Shannon distance between per-VPN destination-country distributions, for each source country. Values near 0 indicate agreement on where endpoints are contacted; values near 1 indicate divergence. Pockets of high divergence (CL, AR, SE, IT, FR) show that VPN choice can substantially reshape the observed destination graph. 
    }
    \label{fig:jsd_heatmaps}
\end{figure*}